\documentclass[twocolumn,showpacs,amsmath,amssymb,aps,prl]{revtex4}

\newcommand{\ben}{\begin{eqnarray}}
\newcommand{\een}{\end{eqnarray}}

\newcommand{\be}{\begin{equation}}
\newcommand{\ee}{\end{equation}}

\newcommand{\nd}{{\noindent}}
\usepackage{color}

\usepackage{graphicx}
\usepackage{amsmath}

\begin{document}

\title{Pure-state density matrix  that competently describes classical chaos}

\author{A.M. Kowalski, A.Plastino and G. Gonzalez Acosta}
\affiliation{Departamento de F\'{\i}sica-IFLP,
Universidad Nacional de La Plata, C.C. 67, La Plata (1900), Argentina}

\begin{abstract}
We work with reference to a well-known semiclassical model, in which quantum degrees of freedom interact with classical ones. We show that, in the classical limit, it is possible to represent classical results (e.g., classical chaos) by means a pure-state  density matrix.
\end{abstract}

\maketitle

\section{Introduction}
\label{sec:intro}

\nd The classical-quantum transition and the  classical limit are certainly frontier issues that constitute an important physics topic \cite{Halliwell,Everitt,Zeh1999,Zurek1981,Zurek2003}. On the other hand, the use of semi-classical systems to describe problems in physics has a long
 historical \cite{Bloch,Milonni,Ring}. A particularly important case is to be highlighted, in which quantum effects in one of the two components of a composite system are negligible in comparison to those in the other. Regarding this scenario as classical simplifies the description and provides deep insight into the combined system dynamics \cite{RK1,RK2,RK3}. This methodology is widely used for the interaction of matter with a field. In this effort we will look at these matters through  a well-known semi-quantum model \cite{Cooper1998,Kowalski2002}. This model
 has been  investigated in great detail  from a purely dynamic viewpoint \cite{Kowalski2002,K0,K1} and also using statistical quantifiers derived from Information Theory (IT) \cite{BP,Tsallis2,Judge,RelativeTsallis}.
 For this model and in \cite{previo}, a suitable density matrix was found for describing the system's route on its way to the classical limit. Rather exhaustive numerical results were presented.

\nd The purpose of this work is to  analytically  determine what happens with the above mentioned pure-state density matrices in the exact classical limit. Same interesting insight will ensue.

\section{Model}
\label{model}

\nd We will consider a Hamiltonian $\hat{H}$ containing classical degrees of freedom (DOF) interacting with strictly quantum DOFs. The dynamical equations for the quantum operators will be the canonical ones \cite{K0,K1},
i.e., any operator $O$ evolves (in the Heisenberg picture)  as
\begin{equation}
  \frac{d O }{dt} = - i\hbar [\: H,  O \:]\,.
  \label{Eccanoncero}
\end{equation} The concomitant evolution equation for its  mean value
$\langle O\rangle\equiv {\rm Tr}\,[\rho\,O(t)]$ will be $ \frac{d \langle  O \rangle}{dt} = - i\hbar \langle[\:  H,
  O \:]\rangle$,
where the average is taken with respect to a proper quantum density operator $\rho$.
Additionally,  the classical variables will obey classical Hamiltonian equations of motion, where the generator is  the mean value of the Hamiltonian,  i.e.,
\begin{subequations}
\label{eqclasgen}
\begin{eqnarray}
\frac{dA}{dt} & = & \frac{\partial \langle  H \rangle}{\partial P_A}
\label{ds},\\
\frac{dP_A}{dt} & = & - \frac{\partial \langle  H \rangle}{\partial A}. \label{clasgenb}
\end{eqnarray}
\end{subequations}
The above equations  constitute an autonomous set of coupled first-order ordinary differential equations (ODE). Solving it  allows for a dynamical description in which no quantum rules are violated, i.e., the commutation-relations are trivially conserved for all times. $A$ plays the role of a time-dependent parameter for the quantum system, and the initial conditions are determined by a proper quantum density operator $\hat \rho$.

 \nd We consider now a system representing the zero-th mode
contribution of a strong external field to the production of
charged meson pairs~\cite{Cooper1998,Kowalski2002}, whose
Hamiltonian is
\begin{equation}
\hat{H}~=~\frac{1}{2}\left(~\frac{\hat{p}^{2}}{m_{q}}~+~
                            \frac{{P_{A}}^{2}}{m_{cl}}~+~
                            m_{q}\omega ^{2}\hat{x}^{2}~\right).
\label{H}
\end{equation}
 where  $\hat{x}$ and $\hat{p}$ are quantum operators, while
$A$ and $P_{A}$ are classical canonical conjugate variables. The
term  $\omega^{2}={\omega _{q}}^{2}+e^{2}A^{2}$ is an interaction
one introducing nonlinearity in our problem, with $\omega _{q}$  a frequency.
$m_{q}$ and $m_{cl}$ are quantum and classical masses, respectively. The Hamiltonian (\ref{H}) is a particular case of a family of semiclassical ones, quadratic in  $\hat{x}$ and $\hat{p}$, without linear terms (see below),
This family  has as a time-invariant a quantity  $I$ that relates to the Uncertainty Principle \cite{Kowalski2002} as
\begin{equation}
\label{I}
I~=~\langle \hat{x}^{2}\rangle \langle \hat{p}^{2}\rangle
-\frac{\langle \hat{L}\rangle^{2}}{4} \geq \frac{\hbar^2}{4}.
\end{equation}
$I$  describes the deviation of the semiquantum system from the classical one given by $I=0$. The quantity $\hat{L}$ is defined as $\hat{L}= \hat{x} \hat{p}+ \hat{p} \hat{x}$.  To investigate  the classical limit one needs also  to consider the classical counterpart of the Hamiltonian (\ref{H}), in which all  variables are classical. In this case $\hat{L}$ is replaced by $L= 2 x p$.
 \nd
We analyze in this work the limit $I\rightarrow0$. A well known ODE-theorem  establishes uniqueness and a continuous dependence of the ODE-solutions on the initial conditions,  if a condition called the Lipschitz one is fulfilled \cite{ODE}.  If the ODE solutions remain bounded as time grows towards infinity, the condition is always satisfied.

\nd Consider semiquantum systems (SS) governed by operators that close a partial Lie algebra with the Hamiltonian. These SS' dynamics will be ruled by closed systems of equations (CSE), involving also the classical variables. These CSE
 will depend in continuous fashion on the initial conditions. For instance, this happens with the set  ($\hat{x}^{2}$, $\hat{p}^{2}$, $\hat{L}$) for quadratic (in $\hat{x}$ and $\hat{p}$) Hamiltonians   \cite{Kowalski2002}. This fact
guarantees the existence of the limit $I \rightarrow 0$ \cite{Kowalski2002}.
%andres
 If the Hamiltonian includes lineal terms in  $\hat{x}$ and $\hat{p}$, $I$ is no longer a constant of the motion. In this case one usesof
$I_{\Delta} = \Delta^{2} x  \, \, \Delta^{2} p  -  \frac{\Delta L^2}{4}$,  which is a time-invariant quantity, instead $I$. The pertinent analysis is similar to the one above described.

\section{MaxEnt Density operator for the semiquantum problem}

\nd We assume \begin{itemize} \item complete knowledge about the initial conditions of the classical variables
\item  incomplete knowledge regarding  the system's quantum components. \item  We only know the initial values of the quantum expectation values of the set 
 of operators $\hat O_{1} =\hat{x}^{2}$, $\hat O_{2} = \hat{p}^{2}$, $\hat O_{3} = \hat{L}$.
\item This set is the smallest  one that carries information regarding the uncertainty principle (via $I$).
\end{itemize}
The MaxEnt statistical operator $\hat \rho$ is given by  \cite{previo}
\begin{equation}
\label{rholevel1}
\hat \rho = \exp{ -\left(\lambda_{0}  \mathcal{ \hat I} + \lambda_1  \hat{x}^{2} + \lambda_2 \hat{p}^{2} + \lambda_3 \hat{L} \right)},
\end{equation}
where the Lagrange multipliers $\lambda_{i}$ are determined
so as to fulfill the set of constraints posed by our prior information
(i.e., normalization of $\hat \rho$ and the supposedly a priori known EV's)
\begin{equation}
\label{constraints}
\langle \hat O_{i} \rangle = {\rm Tr} \;[ \; \hat \rho \;  \hat O_{i}
\; ] \; , \hskip 1.0cm i = 0, \ldots, 3,
\end{equation}
\noindent
($\hat O_{0}=\mathcal{ \hat I}$ is the identity operator). A simplified way to obtaining the values of the multipliers is that of solving the coupled set of
equations \cite{Katz}
\begin{equation}
\label{katze}
\frac {\partial \lambda_0}{\partial \lambda_i}\,=\,-\,
 \langle \hat O_{i} \rangle, \hskip 1.0cm i = 1, 2, 3,
\end{equation}
where
\begin{equation}
\label{landa0}
\lambda_{0}= {\rm Tr}\left[ \exp \left(-\sum_{i=1}^{3}\lambda _{i}\hat O_{i} \right) \right].
\end{equation}
  Using  Eq. (\ref{katze}), one can determine   the ``initial'' $\hat \rho$ given by (\ref{rholevel1}). On the other hand, the statistical operator must
evolve in time from (\ref{rholevel1}) according to the Lioville-von Neumann equation
\begin{equation}
\label{Liouville}
i\hbar \frac{d \hat \rho}{dt}(t) = [\; \hat H, \hat \rho(t) \;]~.
\end{equation}
\noindent
 As the operators $\hat O_{i}$ close a partial Lie algebra with
respect to the Hamiltonian  $\hat H$  \cite{Katz,Levine}, we have
\begin{equation}
\label{conm}
[\: \hat H(t), \hat O_{i} \:] = i\hbar \sum_{j=1}^{3}g_{ji}(t) \hat
O_{j} \;,
\hskip 1.0cm i = 0, 1, \ldots, 3,
\end{equation}
 the statistical operator depends on the time $t$ according to \cite{Levine}
\begin{equation}
\label{rholevelt}
\hat \rho(t) = \exp{ -\left(\lambda_{0} \hat
I + \lambda_1 (t) \hat{x}^{2} + \lambda_2 (t) \hat{p}^{2} + \lambda_3  (t)\hat{L} \right)}.
\end{equation}
provided that the Lagrange multipliers $\lambda_{j} (t)$  verify the set of differential
equations \cite{Levine}
\begin{equation}
\label{lambda}
\frac{d \lambda _{i}}{dt}(t) =  \sum_{j=1}^{3} g_{ij} \lambda _{j}(t) \;,
 \hskip 1.0cm i = 1, 2, 3,
\end{equation}
\noindent
with $\lambda_{j} (0)=\lambda_{j} $ from (\ref{rholevel1}).  The demonstration of this property can be
encountered in the celebrated article \cite{Levine} and is based on the uniqueness of the solutions of the Liouville Equation and the MaxEnt principle, together with the conservation of the Entropy
\begin{equation}
\label{S}
S(\hat \rho) =-{\rm Tr}\;[\; \hat \rho  \;  \ln \hat \rho \;] \; = \lambda_{0}
 + \sum_{i=1}^{3}\lambda_{i} \langle \hat O_{i} \rangle \;,
\end{equation}
\noindent
(Boltzmann's constant is set equal to unity), which is  maximized by the statistical operator (\ref{rholevelt}).

 \nd From now on we will use the fact  that $\lambda_{j} (t)=\lambda_{j} $ to simplify the notation.
In this way, Eqs. (\ref{rholevel1})--(\ref{landa0}) are valid for all $ t $.
Additionally, once  $\hat \rho(t))$ is obtained, we can determine (in the Schr\"{o}dinger picture), \textbf{the temporal evolution of the EV
 of any operator $\hat O$ through}
\begin{equation}
\label{mean values}
\langle \hat O \rangle (t) = {\rm Tr} [\hat \rho(t) \hat O].
\end{equation}

%%%%%%%%%%%%%%%%%%%%

  \nd Note that in this type of semiclassical problem, the $g_{ij}$ of Eqs. (\ref{conm}) and (\ref{lambda}) depend on the classical variables $A$ and $P_A$.
 We use equation (\ref{mean values}) (with $\hat O=\hat H$) in order to
obtain $\langle \hat H \rangle $ and thus describe, via Eqs.
 (\ref{eqclasgen}),  the temporal evolution of $A$ and $P_A$.
The idea is then to regard the set  of equations (\ref{lambda}), together
with the equations (\ref{eqclasgen}), as a single autonomous first-order system.
 Note that the classical equations in turn depend on the mean values. In this case
 the presence of the term  $\langle \hat{x}^{2} \rangle$ in the equation for $P_A$ introduces  an additional  non-linearity
(as such a term is a function of the multipliers)
through
\begin{equation}
\label{O4,L}
\langle \hat{x}^{2} \rangle (t)= {\rm Tr} [\hat \rho(t)
\hat{x}^{2} ],
\end{equation}
\noindent
but we will presently see that this non-linearity can be easily  handled.

\section{Some convenient mathematical results}

\nd It is necessary to calculate $\lambda_{0}$ to relate the initial values of the multipliers and their respective EV's, using Eq. (\ref{katze}).
 We begin by performing  a  change of representation,  made by recourse to the unitary transformation  \cite{previo}
\begin{subequations}
\label{Trans}
\begin{eqnarray}
\hat x & = & \frac{\sqrt{2}}{2} \left( \frac{\lambda_2} {\lambda_1}\right)^{1/4}\left( \left( \frac{\lambda_T} {\lambda_V}\right)^{1/4} \hat X + \left( \frac{\lambda_V} {\lambda_T}\right)^{1/4} \hat P \right), \\
\hat p & = &  \frac{\sqrt{2}}{2} \left(\frac{\lambda_1} {\lambda_2}\right)^{1/4}\left(- \left( \frac{\lambda_T} {\lambda_V}\right)^{1/4} \hat X + \left( \frac{\lambda_V} {\lambda_T}\right)^{1/4} \hat P \right),
\end{eqnarray}
\end{subequations}
where $\lambda_V = \sqrt{\lambda_1 \lambda_2} + \lambda_3$ and $\lambda_T = \sqrt{\lambda_1 \lambda_2} - \lambda_3$.
For reasons of convergence, $\lambda_1 $, $\lambda_2 $, and $\lambda_{1} \lambda_{2} - {\lambda_{3}}^{2} $ must be positive. Then, $\lambda_V$ and $\lambda_T$ become positive too and $I_{\lambda}$ in (\ref{Ilambda}) is well defined.
Of course, the transformation (\ref{Trans}) preserves commutation relations. Thus,  $I$ is also preserved. These new operators are not dimensionless ones [they are expressed in units of the square root of an action and do not depend on $\hbar$,
which is a convenient fact at the time of going over to thhe classical limit].  Further, $\hat X$ and $\hat P$, via
the $\lambda$'s that appear as coefficients in their definition,  are explicitly time-dependent and contain all the relevant information regarding the classical degrees of freedom.
Now $\rho(t)$ becomes \cite{previo}
\begin{equation}
\label{rholevel1I}
\hat \rho(t) = \exp( -\lambda_{0}) \exp \left[-  I_\lambda  \left( \hat{X}^{2} +  \hat{P}^{2} \right) \right].
\end{equation}
\nd The quantity $I_{\lambda}$ defined as
\begin{equation}
I_{\lambda}  =  {\left( \lambda_{1} \lambda_{2} - {\lambda_{3}}^{2} \right) }^{1/2},
\label{Ilambda}
\end{equation}
\noindent
 a constant of the motion \cite{previo}. This invariant is the equivalent of the one in  Eq. (\ref{I}), expressed in terms of the
 $\lambda$'s.

\nd Despite the characteristics assigned to  $\hat X$ and $\hat P$, the operator  $\hat{X}^{2} +  \hat{P}^{2}$ has a discrete spectrum, one resembling that of a the Harmonic Oscillator, because the commutation relations are preserved for all time.
 After a little algebra, it is easy to see from (\ref{landa0})
that
\begin{equation}
\label{landa0parti}
\lambda_{0}= - \ln \left[\exp (\hbar \, I_{\lambda})- \exp (-\hbar \, I_{\lambda}) \right],
\end{equation}
and using Eq. (\ref{katze}) (or Eq. \ref{mean values}), the particular EV's can be cast in the fashion \cite{previo}
\begin{subequations}
\label{rel.mv,L}
\begin{eqnarray}
\langle \hat{x}^{2}\rangle & = &
  \frac{T(I_{\lambda})}{I_{\lambda}} \lambda_{2}, \label{rel.mv4,xcuad} \\
\langle \hat{p}^{2}\rangle & = &
\frac{T(I_{\lambda})}{I_{\lambda}} \lambda_{1}, \\
\langle \hat{L} \rangle & =  & -2
\frac{T(I_{\lambda})}{I_{\lambda}} \lambda_{3}, \label{rel.mv4,L4}
\end{eqnarray}
\end{subequations}
\noindent
with $T(I_{\lambda})$ given by \cite{previo}
\begin{equation}
\label{tilambda}
T(I_{\lambda})  =  \frac{\hbar}{2}\,\, \left(\frac{\exp(2 \, \hbar \, I_{\lambda})+1}
{\exp(2 \,\hbar\, I_{\lambda})-1}  \right).
\end{equation}
Further, we deduce from (\ref{rel.mv,L}) that
\begin{equation}
\label{I, T(Ilamba)}
T(I_{\lambda}) \,= \,\sqrt{I \,},
\end{equation}
 Now, by recourse to the Eqs. (\ref{eqclasgen}), (\ref{lambda}), and (\ref{rel.mv4,xcuad}), we are in position to write down our dynamical system of equations as a closed one in both multipliers and classical variables. We have  \cite{previo}
\begin{subequations}
\label{ODE}
\begin{eqnarray}
\frac{d\lambda_{1}}{dt} & = & 2 m_{q} \omega^2 \lambda_{3},\\
\frac{d\lambda_{2}}{dt} & = & - \frac{2}{m_{q}}\lambda_{3},\\
\frac{d\lambda_{3}}{dt} & = & - \frac{1}{m_{q}}\lambda_{1} + m_{q} \omega^2 \lambda_{2},\\
\frac{dA}{dt} & = & \frac{P_{A}}{m_{cl}}, \\
\frac{dP_{A}}{dt} & = & -e^{2}m_{q}\,A \frac{T(I_{\lambda})}{I_{\lambda}} \lambda_{2}. \label{lamb+s,p,p}
\end{eqnarray}
\end{subequations}
\noindent
  This system associates \textbf{a kind of phase-space} to
the density operator (\ref{rholevelt}), determined by classical variables and Lagrange multipliers. The system (\ref{ODE}) depends in nonlinear fashion upon the classical
variable  $A$, via $\omega^2$, but the non-linear term $T(I_{\lambda})$ in
(\ref{lamb+s,p,p}) is easily tractable as a  function of $I$, using (\ref{I, T(Ilamba)}). This non-linearity is thus replaced by a
dependence upon $I$ plus the initial conditions. This last dependence emerges via the
invariant $I_{\lambda}$ (which in turn is fixed by $\hat \rho(0)$, i.e. by the
initial values of the Lagrange multipliers).

\section{Useful previous results}

 \nd In \cite{previo}, we investigated the dynamics described by the density operator (\ref{rholevelt})  as a function of the
relative  energy $E_r$, defined as $E_{r}=\frac{|E|}{I^{1/2} \omega _{q}}$. The classical limit obtains for  $E_r \rightarrow \infty$ (a particular case is $I \rightarrow 0$, which we will  study below).

\nd In \cite{previo} we also showed that, by augmenting  $E_r$ (for example decreasing $I$), the physical system passes through  three regions: a quasiclassical one, a transitional one,  and a classical one. As  $E_r$ grows, complexity augments and, eventually, chaos emerges. This is a phenomenon of a semi-classical nature, since the classical dynamics-stage has, obviously, not yet been reached. Remark on  the coexistence of the Uncertainty Principle with chaos and also on that, having  $\hat \rho (t)$, one can know the time dependence of any expectation value  via Eq. (\ref{mean values}).

Also, from Eqs. (\ref{tilambda}) and (\ref{I, T(Ilamba)}) we found in  \cite{previo} that
\begin{equation}
\label{I-Ilamba}
I_{\lambda} \,= \,\frac{1}{2\,\hbar} \ln \left( \frac{\sqrt{I}+\frac{\hbar}{2}}{\sqrt{I}-\frac{\hbar}{2}} \right),
\end{equation}
relating $I_{\lambda} $ to $I$.
Note here that as $I$ decreases,  $I_{\lambda}$ augments. If $I$ approaches $\hbar^{2}/4$, then $I_{\lambda} \rightarrow \infty$, since ${X}^{2} +  \hat{P}^{2}$ approaches the ground state. Even then  $I\neq0$. Thus, we do not reach the classical limit yet.
 We need to take the limit  $\hbar \rightarrow 0$ and still   $I_{\lambda} \rightarrow \infty$ holds \cite{previo}.

\section{Present results regarding the  classical limit (CL)}

\nd Our present elaborations begin at this point. We are going to analytically  study the limit $ I \rightarrow 0 $ of the density operator (\ref{rholevel1I}). Speaking of a CL entails  that both  $\hbar$  and $I$ $\rightarrow 0$, even if  our Evs numerical results are independent of the actual numerical value of $\hbar$. In going to this limit we must always respect the restriction (\ref{I}). Two roads are open to us
\begin{enumerate}
\item Take first $\hbar\rightarrow 0$ and then  $I\rightarrow 0$. Classical statistics and quantum one  are both compatible with  (\ref{I}), for any  $\hbar>0$ (quantum) or for $\hbar=0$ (classic). In the limit $\hbar\rightarrow 0$, the density matrix  (\ref{rholevel1I}) adopts the form
\begin{equation}
\label{Imixta}
\rho= \frac{\mathcal{I}}{Tr[\mathcal{I}]},
\end{equation}
with $\mathcal{I}$ the identity matrix.  One has

\begin{equation}
\label{I-Ilamba cl lim}
\lim\limits_{\hbar \rightarrow 0} I_{\lambda} = \frac{1}{2 \,\sqrt{I}},
\end{equation} as a result of
\begin{equation}
\label{0}
\lim\limits_{\hbar \rightarrow 0}\, \hbar \, I_{\lambda}=0,
\end{equation}
where we employed Eq. (\ref{I-Ilamba}).  (\ref{Imixta}) is the maximally mixed density matrix of diagonal  elements  $1/n, n \,\varepsilon \,\mathbb{N} $, with  $n\rightarrow \infty$. Such matrix should arise out of a decoherence process.
 We can not now take the limit  $I$ $\rightarrow 0$.

\item  Proceed to effect  $ \lim\limits_{\hbar \rightarrow 0}\, \lim\limits_{I \rightarrow \hbar^{2}/4} \Lambda$,  $\Lambda$   referring here to any of our quantities of interest. This second choice of venue respects the restriction (\ref{I}) and would constitute the correct way to go. According to (\ref{I-Ilamba}), we have
\begin{subequations}
\label{infty}
\begin{eqnarray}
\lim\limits_{\hbar \rightarrow 0}\, (\lim\limits_{I \rightarrow \hbar^{2}/4} I_{\lambda})&=& \infty,\\
\lim\limits_{\hbar \rightarrow 0}\, (\lim\limits_{I \rightarrow \hbar^{2}/4}\hbar \, I_{\lambda})&=& \infty,\\
\lim\limits_{\hbar \rightarrow 0}\, (\lim\limits_{I \rightarrow \hbar^{2}/4} \lambda_i)= \infty, & & i = 0, 1,2, \\
\lim\limits_{\hbar \rightarrow 0}\, (\lim\limits_{I \rightarrow \hbar^{2}/4} | \lambda_3|)&=& \infty.
\end{eqnarray}
\end{subequations}
\end{enumerate}

\nd Note that in the second instance, when  $I$ tends to its minimum possible value $\hbar^{2}/4$, $\rho$  (\ref{rholevel1I}) tends to its ground state. Thus, considering the pseudo \textit{generalized temperature}  $1/I_{\lambda}$, we ascertain that
  $1/ I_{\lambda}\rightarrow 0$. Remark that  $I_{\lambda}$ depends on both the classical variables and the initial conditions for the  EVs. Our results holds also for $\hbar \rightarrow 0$. Lo and behold, we have found that the classical limit is represented by a pure-state density matrix!.
	
\nd Looking at the asymptotic behavior of 	$\lambda_0$ en (\ref{landa0parti}), we see that $\exp (-\lambda_{0}) \sim  \exp (\hbar \, I_{\lambda})$, entailing that the asymptotic eigenvalues of $\rho$ become $\exp{[-n \hbar I_\lambda]}$, $n=0, 1, 2,  \ldots $.   Thus,  $\rho$ (\ref{rholevel1I}) (or (\ref{rholevel1})), asymptotically, in its eigen-basis  has the associate  density matrix ${\cal R} (t)$ 
\begin{equation}
{\cal R}(t)=\left(\begin{array}{cc}
     1 \,\,\, 0 \,\,\, 0 \,\,\,\ldots \\
     0 \,\,\, 0 \,\,\, 0 \,\,\,\ldots \\
     0 \,\,\, 0 \,\,\, 0 \,\,\,\ldots \\
     \mathbf{\vdots}
\end{array}\right).
\label{Ademo}
\end{equation}
This is a rather surprising. Not only the classical features of the semiclassical evolution depicted in Figs. 1 and 2 of \cite{previo} are represented by a mixed quantum density matrix, but purely classical results with $I=0$, are masked by a pure-state density matrix. In the first case, semi-classical Chaos is obtained. In the second, directly classical Chaos.

Any asymptotic mean value will evolve \textbf{classically}. Evs $ \langle \hat {X}^{n} \hat{P}^{m} \rangle $ will be null for all time, thus being trivially classic.

Additionally, not only the mean values of the set ($\hat{x}^{2}$, $\hat{p}^{2}$, $\hat{L}$) will evolve asymptotically with the classical equations corresponding to the Hamiltonian totally classic, if not surprisingly enough, $ \langle \hat {x}^{n} \hat {p}^{m} \rangle(t) = \langle \hat {x} \rangle^{n} \langle \hat {p} \rangle^{m} (t) $ with initial conditions deducible as powers of
 $ \langle \hat{x}\rangle (0)$ and $\langle \hat{p}\rangle (0)$, with $ \langle \hat{x}\rangle (0)= \pm\sqrt{\langle\hat{x}^{2}\rangle (0) }$ and $ \langle \hat{x}\rangle (0)= \pm\sqrt{\langle\hat{p}^{2}\rangle (0) }$.
This follows from (\ref{mean values}) and ($\ref{Trans}$)  after slight manipulation.

\nd  As a proof of the correctness of our results, it is easy to see that $I$  calculated with $\rho(t)$  given by  (\ref{Ademo}) vanishes. Denoting the ground state by $|0>$, we have $ <0|\hat X^2|0>=<0|\hat P^2|0>= \lim\limits_{\hbar \rightarrow 0} \hbar/2$ and $<0|\hat L|0>=0$, so that  $I=0$.
 Moreover,  via (\ref{S}), we obtain an entropy  $S= - \lambda_0 - 2 \, I_{\lambda} \sqrt{I}$, a decreasing monotonic function of $I$, with asymptotic value $S=0$, as expected for a pure state.

\subsection{The pertinent classical statistical treatment}
\label{estadisticaclasica}

\nd Let us think of
\begin{equation}
\label{rhoclasica}
 \rho(x,p,t) = \exp -\left(\lambda_{0cl} + \lambda_{1cl}  {x}^{2} + \lambda_{2cl} {p}^{2} + \lambda_{3cl} {L} \right),
\end{equation}
equivalent to  (\ref{rholevelt}). Here  ${x}^{2}$, ${p}^{2}$ and  $L=2xp$ are simple functions, of course. The mean value of any general $F(x,p,t)$, for all  $t$, is given via  $ \int_{-\infty}^{\,\infty} \int_{-\infty}^{\,\infty} F(x,p,t)\, \rho(x,p,t) \, dx dy$.  Using a transformation equivalent to (\ref{Trans}), but for classical variables, we obtain the classical version of (\ref{rholevel1I}), with $\lambda_{0 cl}= \ln (\pi / I_{\lambda cl})$.
After some manipulation we are led to
\begin{subequations}
\label{rel.mv,Lcl}
\begin{eqnarray}
\langle {x}^{2}\rangle & = &
  \frac{\sqrt{I_{cl}}}{I_{\lambda cl}}\, \lambda_{2 cl},  \\
\langle {p}^{2}\rangle & = &
\frac{\sqrt{I}_{cl}}{I_{\lambda cl}}\, \lambda_{1 cl},  \\
\langle L \rangle & =  & -
\frac{ 2 \, \sqrt{I_{cl}}}{ I_{\lambda cl}} \, \lambda_{3 cl}.
\end{eqnarray}
\end{subequations}
\noindent
Now $I_{\lambda cl}$ is the classical version of  (\ref{Ilambda}),
\begin{equation}
I_{\lambda cl}  =  {\left( \lambda_{1 cl} \lambda_{2 cl} - {\lambda_{3 cl}}^{2} \right) }^{1/2},
\label{Ilambdacl}
\end{equation}
\noindent
 that is also a time-invariant quantity, since the $\lambda_{i cl}$ obey  the same system of equations used in the quantum treatment (Eqs. \ref{ODE}). The classical version of  $I$  reads
 $I_{cl}~=~\langle {x}^{2}\rangle \langle {p}^{2}\rangle
- \langle {L}\rangle^{2}/4 $,
 but it no longer satifies  (\ref{I}), but obeys instead
 \noindent
\begin{equation}
\label{Icl}
I_{cl} \geq 0.
\end{equation}
 Moreover,  Eqs. (\ref{rel.mv,Lcl}) coincide with Eqs. (\ref{rel.mv,L}), together with (\ref{I, T(Ilamba)}).
However, in this instance the dependence of  $I_{\lambda cl}$ with  $I_{cl}$ is not given  by  (\ref{I-Ilamba}), since
\begin{equation}
\label{I-Ilamba cl}
I_{\lambda cl} = \frac{1}{2 \,\sqrt{I_{cl}}},
\end{equation}
but will coincide with  Eq. (\ref{I-Ilamba cl lim}), as one may expect. These classical results  confirm that the limit $\hbar \rightarrow 0$ is consistent with both  classical and  quantum statistics.
Obviously, to complete the present  analysis,  the limit $I_{cl} \rightarrow 0$ (or $I_{\lambda cl} \rightarrow \infty$) is demanded.
 The probability density function (\ref{rhoclasica}) will read
\begin{equation}
\label{rhoclasicalimite}
\lim\limits_{I_{cl} \rightarrow 0}  \rho(x,p,t) = \delta(X) \delta(P),
\end{equation}
being a Dirac delta function of $X$ and $P$, as one should expect. In the  limit (\ref{rhoclasicalimite}), also $\langle \hat{X}^{n}\hat{P}^{m} \rangle =0$ at all times and all results with total certainty are obtained via ($\ref{Trans}$). Total certitude is achieved without need for any kind of statistical reasoning. 

\section{Conclusions}
\label{sec:Concl}

In this work we have exhaustively investigated the classical limit of a density operator  $\rho$ associated to a well-known non-linear semi-classical system  that possesses both  classical and quantum interacting degrees of freedom. This  $\rho$
was presented previously in \cite{previo}, in a context of {\it incomplete} prior information.

\nd In \cite{previo} its authors detected three well delimited and different regions in traversing the road towards the classical limit. These zones were characterized by the parameter $E_{r}=\frac{|E|}{I^{1/2} \omega _{q}}$, con $E_r \rightarrow \infty$, with $E$ the total energy and $I$ a dynamical invariant intimately linked to the uncertainty principle.  

\nd One had a quasiclassical region, a transitional one,  and a classical zone. As  $E_r$ grows, complexity augments and, eventually, chaos emerges. This was a phenomenon of a semi-classical nature.

\nd {\it It is article focused attention specifically  on the classical limit per se, not on the road to it as in  \cite{previo}}.

\nd A purely analytical treatment was  effected, for $I \rightarrow 0$. Two possible paths were contemplated to perform our study.
The first was to research the $\hbar\rightarrow 0$ calcculation. Some difficulties were encountered in such instance, that were discussed in the text.

\nd The second path turned to be both  correct and coherent. It consist in taking first   $\lim I \rightarrow \hbar^{2}/4$, approaching the minimum $I-$value that quantum mechanics permits. A posteriori one deals with the limit $\hbar \rightarrow 0$.
In quite a counter-intuitive fashion, we stumbled on an asymptotic density matrix ${\cal R}$ corresponding to a pure state (\ref{Ademo}). ${\cal R}$ adequately describes all facets of our classical features. i.e., those pertaining
to  a classical Hamiltonian. Any and all ${\cal R}$-mean values  behave as classical variables. For example,   $\langle \hat{x}^{n}\hat{p}^{m} \rangle (t) =\langle \hat{x}\rangle^{n} \langle \hat{p}\rangle^{m} (t)$.  We conclusively showed that
${\cal R}$ competently describes classical chaos.

\section{Acknowledgments}

A. M. K. fully acknowledges support from the Comisi\'on  de Investigaciones Cient\'{\i}ficas de la
Provincia de Buenos Aires (CICPBA) of Argentina.

\end{document}